\documentclass[smallextended]{svjour3}

\usepackage{epstopdf}
\usepackage[caption=false]{subfig}

\usepackage[numbers,sort&compress]{natbib}
\usepackage{xcolor}
\usepackage{algorithm}
\usepackage{algpseudocode}
\usepackage[title]{appendix}
\usepackage{dsfont}
\usepackage{booktabs}
\usepackage{colortbl}
\usepackage{caption}
\usepackage{bm}
\usepackage{graphicx}
\usepackage{amsfonts}
\usepackage{amsmath}

\usepackage{wrapfig}
\usepackage{lscape}
\usepackage{rotating}
\usepackage{epstopdf}

\bibpunct[, ]{[}{]}{,}{n}{,}{,}

\newtheorem{prop}{Proposition}[section]
\newtheorem{assu}{Assumption}[section]
\definecolor{intnull}{RGB}{213,229,255}

\begin{document}

\title{Modeling Dependence via Copula of Functionals of Fourier Coefficients}

\author{Charles Fontaine \thanks{Corresponding author. e-mail: charles.fontaine@kaust.edu.sa} \and Ron D. Frostig \and Hernando Ombao}

\institute{C. Fontaine \and H. Ombao \at
              Statistics Program, King Abdullah University of Science and Technology (KAUST),
							23955 Thuwal (Saudi Arabia)     
           \and
           R. D. Frostig \at
					  Departments of Neurobiology and Behavior, Biomedical Engineering, and the Center for Neurobiology of Learning and Memory, University of California-Irvine, Irvine, CA, 92697 U.S.A.}
					
\maketitle

\begin{abstract}

The goal of this paper is to develop a measure for characterizing complex dependence  between stationary time series that cannot be captured by traditional measures such as correlation and coherence. Our approach is to use copula models of functionals of the Fourier coefficients which is a generalization of coherence. Here, we use standard parametric copula models with a single parameter both from elliptical and Archimedean families. Our approach is to analyze changes in local field potentials in the rat cortex prior to and immediately following the onset of stroke. We present the necessary theoretical background, the multivariate models and an illustration of our methodology on these local field potential data. Simulations with non-linear dependent data show information that were missed by not taking into account dependence on specific frequencies. Moreover, these simulations demonstrate how our proposed method captures more complex non-linear dependence between time series. Finally, we illustrate our copula-based approach in the analysis of local field potentials of rats. 

\end{abstract}

\noindent%
\vfill

\section{Introduction}
\label{sec:intro}

Consider an experimental setting where multichannel brain signals are recorded continuously from an animal (rat, monkey, human) over a certain period of time. The key scientific questions being addressed often center 
on brain connectivity, that is, how different brain regions interact. In particular, the emphasis on these finely-sampled brain electrophysiological signals (e.g., local field potentials (LFP)) is on interactions between oscillatory components extracted from each channels. Methods for 
analyzing dependence between brain signals have been developed in the 
literature (see, e.g., \cite{FiecasPCoh}
and \cite{Ombaobook}) and a 
more formal and general treatment of spectral analysis is discussed in 
\cite{ShumwayStoffer}. However, classical spectral metrics 
(e.g., coherence and partial coherence) are limited in that they can 
capture only the strength of the linear dependence between the Fourier coefficients. The goal of this paper is to develop a rigorous approach that can comprehensively model general dependence structures between oscillatory activity of time series via copulas but using spectral features such as functionals of the Fourier coefficients.    

\begin{figure}
\begin{center}
\includegraphics[scale=0.40 ]{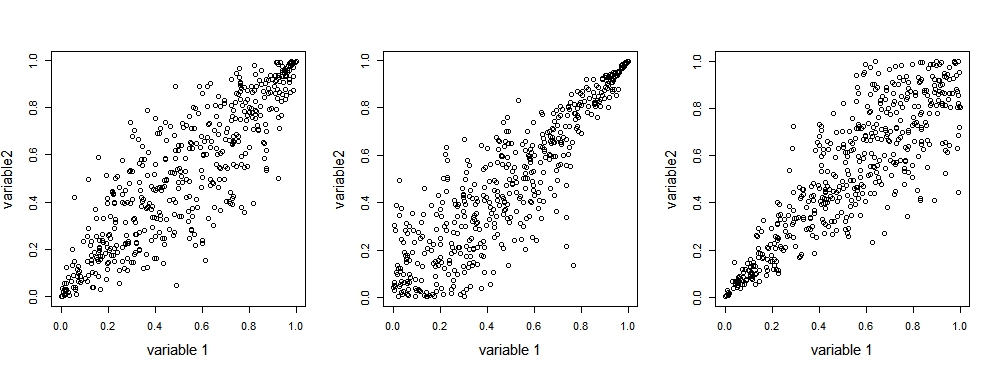}
\caption{Illustration of cases of non-linear dependence \textit{(middle and right)} which cannot be detected by classical measures. All cases have a similar Kendall's tau measure of $0.7$, but in the middle figure, there is strong dependence between two variables when these are at the upper tails. In the figure on the right, that strong dependence is when these are at the lower tail.}
\label{exampleIntro1}
\end{center}
\end{figure}

In Figure \ref{exampleIntro1}, one observes three dependence structure having the same correlation measure (or value). The difference between these cases cannot be captured by a linear dependence measure (e.g. correlation, partial correlation, coherence, partial coherence). Hence, our goal in this paper is to present a copula-based framework to deal with these complex-structured dependencies in the spectral domain, for particular fundamental Fourier frequencies. We present a semi-parametric copula-based methodology to express these complex dependencies. The novelty here is that we develop a new approach that incorporates major statistical features (Kendall's tau, empirical cumulative distribution function) in the semi-parametric copula inference in order to consider spectrally represented data.

We now present an overview of this work. Consider a $d$-dimensional time series segmented into $R$ possibly over-lapping epochs (periods of $1$ second) with $T$ observations within each epoch. We think of this data as signals recorded in $d$ locations on the brain. The time series across all epochs will be assumed to be stationary so that the dependence structure between variables remains constant over the course of all $R$ epochs.  We denote the time series for the $r$-th epoch 
to be the $T \times d$ matrix ${\bf X}^{(r)} = [X_1^{(r)}, \ldots, X_{d}^{(r)}]$ and $X^{(r)}_\ell$, $\ell=1,...,d$, to be the vector $[X_\ell^{(r)}(1),...,X_\ell^{(r)}(T)]'$ corresponding to recordings in channel $\ell$. The time domain approach to model dependence between time series 
is directly via ${\bf X}^{(r)}$. However, if one wants to represent the dependence in the spectral domain between the Fourier fundamental frequency, the main approach is based on \textit{coherence} between channels $X_\ell$ and $X_{\ell '}$, $\ell,\ell '=1,...,d$, at 
frequency $\omega_k$, which is approximately equal to the expected value of the squared 
absolute correlation between the Fourier coefficients. In this paper, we will examine more general (non-linear) dependence between oscillatory components by modeling copulas of functionals of Fourier coefficients.

\begin{figure}
\begin{center}
\includegraphics[scale=0.30 ]{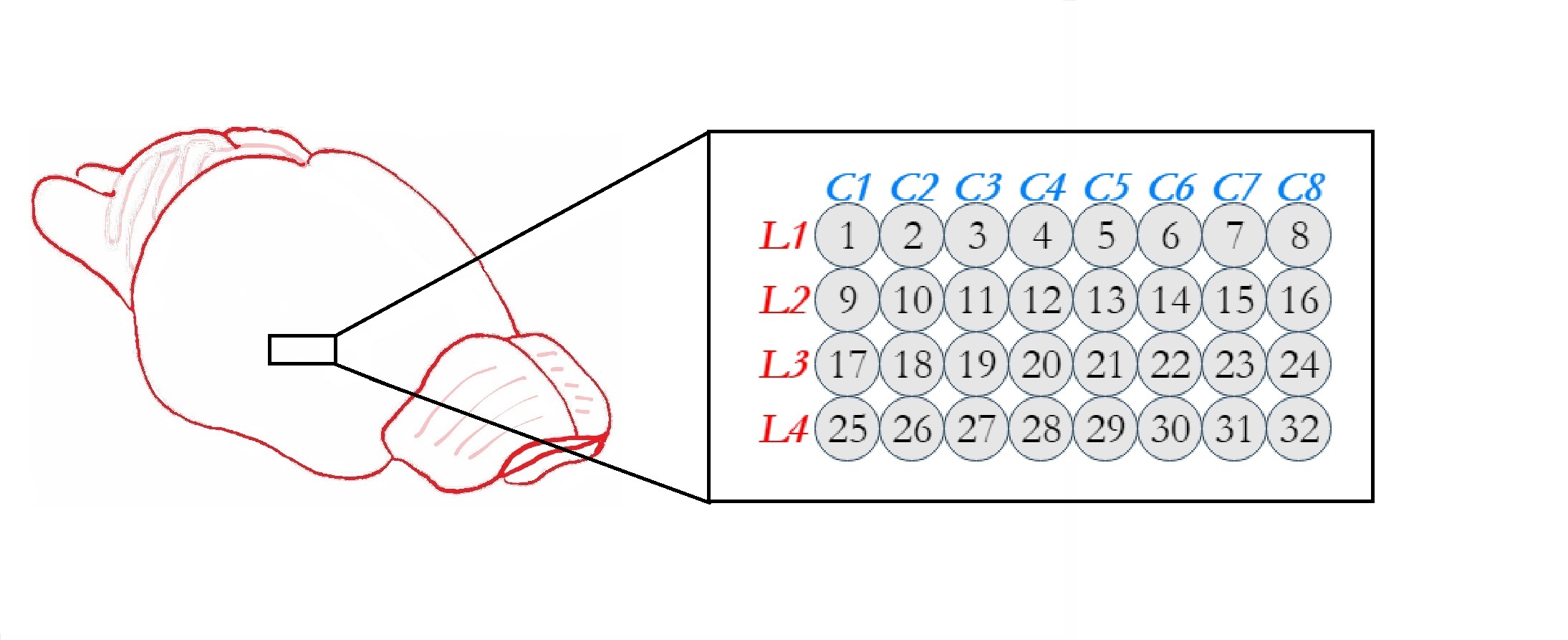}
\caption{Placement of the $32$ microelectrodes in the cortex of the rat. There are $8$ columns (blue) of microelectrodes, each column having $4$ layers (red) that span most of the cortical depth. }
\label{RatBrain}
\end{center}
\end{figure}

The motivation behind this work is the following experiment. At the neurobiology laboratory at the University of California-Irvine (Principal Investigator: Frostig, second author; see \citet{wann2017}), stroke was artificially induced in experimental rat model of ischemic stroke by severing the medial cerebral artery (MCA). Brain activity prior to and after the stroke was recorded through the local field potentials (LFPs) from $d=32$ microelectrodes placed directly within the rat cortex. The recording time window covered $5$ minutes pre-stroke and $5$ minutes post-stroke. For each second, $T=1000$ time points were recorded and analyzed.  Figure \ref{RatBrain} shows the placement of the microelectrodes. The key scientific questions being addressed by neuroscientists is focused on stroke-induced changes in brain connectivity (i.e., communication patterns between neuronal populations). In particular, the emphasis on these local field potentials is on interactions between oscillatory components extracted from each microelectrode. Thus, the statistical interest here was to develop a measure that can characterize the complex nature of dependence (particularly in the spectral domain) between the signals recorded by the microelectrodes and to develop a method that can detect changes in dependence following a shock to the brain system (such as a stroke). The experimental setup described above will be detailed in Section \ref{lfp}.

Methods for analyzing dependence between brain signals have been developed in the 
literature (see, e.g., \citet{matousek1973}, \citet{FiecasPCoh}, \citet{Ombaobook}, \citet{guevara1996}, \citet{shaw1981} and \citet{shaw1984}) and a more formal and general treatment of spectral analysis is discussed in \citet{ShumwayStoffer}. However, classical spectral metrics 
(e.g., coherence and partial coherence) are limited in that they can capture only the strength of that linear dependence between the Fourier coefficients. Thus, these measures may miss some complex (or non-linear) dependence structures between two frequencies. For example, taking the well-known work on changes in dependence for brain channels in the spectral domain: \citet{long2004}, \citet{purdon2001}, \citet{nunez2006} and \citet{gotman1982}, one realizes that these authors constrained their work to the linear dependence between signals (or frequencies). Hence, the proposed approach in this paper is an attempt to provide a new tool for capturing more general (beyond linear) dependence between signals.  

We denote the microelectrode $\ell=1,...,d$ and the epoch $r=1,...,R$ with each epoch having $T$ time points (for this specific data, we have $d=32$, $R=600$ and $T=1000$). Thus, the time domain variable representing these microelectrodes on the $r$-th epoch is $X^{(r)}_\ell$. As we are interested to work in the spectral domain in order to capture dependence between oscillatory activity, we compute the Fourier coefficients in order to get the Fourier coefficients  $$f_{\ell, \omega_k}^{(r)} = \frac{1}{\sqrt{T}} \sum_{t=1}^{T} X_\ell^{(r)}(t) 
\exp(-i 2 \pi \omega_k t)$$ where $\omega_k = \frac{k}{T}$, 
\ $k = 0, 1, \ldots, T-1$ \ are the fundamental Fourier 
frequencies. As we work in this paper with the magnitudes of the Fourier coefficients (or square root of periodograms), let $\delta_{\ell,\omega_k}=[\delta_{\ell,\omega_k}^{(1)},...,\delta_{\ell,\omega_k}^{(R)} ]'=[ | f_{\ell,\omega_k}^{(1)}|,...,|f_{\ell,\omega_k}^{(R)}|]'$ being the vector of magnitudes of the Fourier coefficients for microelectrode $\ell=1,...,d$ and for frequency $\omega_k$ over the $R$ existing epochs. Let ${\bm{\delta}}_{\omega_k}=[\delta_{1,\omega_k},...,\delta_{d,\omega_k}]$ to be the matrix of dimension $R \times d$ of these vectors for the $d$ microelectrodes. Let $F_{1,\omega_k},...,F_{d, \omega_k}$ be the marginal distributions associated to ${\bm{\delta}}_{\omega_k}$, admitting a density. Obviously, when components of 
${\bm{\delta}}_{\omega_k}$ are independent, the joint cumulative distribution 
function (cdf) is the simply the product of its marginal cumulative
distributions. However, 
when there is dependence between the different components of the time series, 
\citet{sklar1959} provides an explicit form of the joint cdf $H({\bm{\delta}}_{\omega_k})$  by
\begin{eqnarray} \label{sklar}
H\left( \delta_{1,\omega_k},...,\delta_{d,\omega_k}  \right) &=& C_{1,...,d;\omega_k} \left( \delta_{1,\omega_k},...,\delta_{d,\omega_k}  \right)
\end{eqnarray}
where $C_{1,...,d;\omega_k}$ is a copula: a cumulative distribution function expressing the mapping $[0,1]^d \to [0,1]$. In practice, a copula is characterized by a model which may be either parametric or non-parametric. Under a correct specification, one 
clear advantage of this approach is the flexbility of $C$ in characterizing changes in the nature and strength of 
dependence and yet still retain its general structure. Indeed, if we look at the basic example in Figure \ref{exampleIntro1} to consider both structure and strength, we see the evidence that the dependence between $Variable \text{ }1$ and $Variable\text{ }2$ differs in the three case. Using Pearson's correlation, the equivalent of coherence on real-valued domain, the estimated strength of dependence is $\tau=0.70$, which is not truly reflective of the actual process. The copula model captures the information about that non-linear structure of dependence; in this particular example, there is a distinct difference in the tail dependence. Indeed, one observes a strong linear dependence in the lower tail in the middle figure and a strong linear dependence in the upper tail in the left one.

Copulas have been used to model dependence between random 
variables (e.g., \citet{aas2009} and \citet{joe1997}). However, a 
straightforward application of a standard parametric copula model 
cannot fully capture the essence of certain forms of dependence. Hence, many 
copula-based approaches have been developed to deal with these
complex dependence issues. For example, the empirical multivariate approaches 
(e.g. \citet{deheuvels1979}), the kernel-based approaches 
(e.g. \citet{gijbels1990}) or approaches based on Bernstein 
polynomials approximations (e.g. \citet{li1998} and 
\citet{sancetta2004}) have 
the flexibility to capture more complex dependence. However, while these models are more robust, 
they typically suffer from lower power. 

As a remark, in our work, $C_{\omega_k}$ is assumed to be constant across all epochs $r=1,...,R$ and thus data from all epochs (functionals of Fourier coefficients) will used to estimate this common dependence structure. Some studies have dealt with this frequency-spectral dependence between two or more channels, particularly in an inference perspective. For example, \citet{ibragimov2005} and \citet{lowin2010} theorized the modeling of principle of the Fourier copula, based on the work of \citet{delapena2003}, where the bivariate copula is expressed by $C(u,v)=\int_0^u \int^v_0 (1+g(u,v))dudv,$ for $g(u,v)$ being simply a global 
measure across the entire frequency range. They proposed to estimate this copula by an empirical joint cdf and showed the asymptotic convergence of the related empirical copula process. While seemingly attractive, a major drawback is that we are interested in expressing the dependence between specific frequency bands rather than the entire frequency range. In 
this case, it will not be possible to model the strength of both full and partial dependence through a parameter as we can see in a standard 
elliptic copula using Pearson's correlation matrix as the expression of 
this parameter. 

In this paper, we develop an approach to model dependence while keeping the robustness of parametric copulas
(seeing through the expression of a parameter structure) and using the advantages of the decomposition of time series data in 
band-specific frequency oscillations. The main feature of our methodology is that it uses the coherence or the Kendall rank-based coherence as measures to express the strength of the dependence in a parametric model where margins are expressed from the magnitude of Fourier coefficients. In Section \ref{sec:models}, we present an inferential framework for both elliptical and Archimedean families of copulas. In Section \ref{simuls}, we illustrate the potential of our approach through some simulations on specific cases of idiosyncratic dependencies. Finally, we apply in Section \ref{lfp} the presented methodology on local field potential (LFP) of experimental rats. 

\section{Models}
\label{sec:models}

Any parametric copula consists of three main components: the marginal distributions, the dependence parameter(s) and the copula structure itself. Here, we assume that we use a straightforward copula structure in its classical form. Hence, we have to express both margins and dependence spectrally with regard to its analytic properties, to infer, in order, the cdf and the dependence coefficient(s) - functionals of spectral domain analogy of Pearson or Kendall measures. 

Before presenting our models and inferential framework, we review and set some necessary notation. Let $\delta_{\ell, \omega_k}^{(r)}$ to be the magnitude of the Fourier coefficient $f_{\ell, \omega_k}^{(r)}$ for the microelectrode (brain channel) $\ell=1,...,d$ at frequency $\omega_k=\frac{k}{T},k=0,...,T-1$ for epoch $r=1,...,R$. The collection of all these magnitudes over all the possible epochs is denoted by the vector $\delta_{\ell, \omega_k}=[\delta_{\ell, \omega_k}^{(1)},...,\delta_{\ell, \omega_k}^{(R)} ]'$ and any subvector for $1 \leq r < s \leq R$ will be denoted by $\delta^{(r:s)}_{\ell, \omega_k}=[\delta_{\ell, \omega_k}^{(r)},...,\delta_{\ell, \omega_k}^{(s)} ]'$. Let $\xi_{\ell,\omega_k}$ to be the cumulative distribution function of $\delta_{\ell, \omega_k}$, which admits a density. 

\subsection{Marginals inference}

We note that our goal is to model dependence directly on the magnitude 
of the Fourier coefficients. Thus, we express the distribution of the vector $\delta_{\ell,\omega_k}$ at any
channel $\ell=1,...,d$, by the definition of cdf, i.e.
$\mathbb{P}(\delta_{\ell,\omega_k} \leq y)=\xi_{\ell,\omega_k}(y). $
A natural method to estimate the distribution is to apply 
straightforward the empirical cumulative distribution (ecdf) estimator 
to the components of $\delta_{\ell,\omega_k}$ where the data length is the number of epochs $R$ on which we compute the following
\begin{eqnarray}
\widehat{\xi}_{\ell,\omega_k}(y) &=& \frac{1}{R}\sum_{r=1}^{R} \mathbb{I}(\delta_{\ell,\omega_k}^{(r)} \leq y)
\end{eqnarray}
on a single frequency over all epochs; thus the asymptotic convergence to the real cdf is preserved due to Glivenko-Cantelli (see \citet{vandervaart1998}). One remark is that the latter formula holds only when we are interested by dependency on a particular frequency and that this dependence is assumed to be constant across all epochs. In the situation where we are interested in a frequency \textit{band} (rather than a single frequency) we apply filtering on each epoch and then compute the sum across frequencies in that band, over a single epoch, one should sum on $k\in \mathcal{K}$ where $\mathcal{K}$ is the set containing all possible frequency bands (e.g., the delta frequency band $\mathcal{K}_\Delta=(0,4)$ Hertz). 

\subsection{Dependence parametrization}
We present here two measures of dependence that will serve, using a functional of them, to express the estimation of the dependence parameter of any well-known parametric copula.

\subsubsection{Coherence}
The analogue of the Pearson correlation for the spectral domain 
at frequency $\omega_k$ is coherency which is the ratio of the 
cross-spectrum 
(or covariance between ${\bm{f}}_{\ell,\omega_k}$ and ${\bm{f}}_{\ell',\omega_k}$, $\ell,\ell' =1,...,d$; where ${\bm{f}}_{\ell,\omega_k}=[f_{\ell,\omega_k}^{(1)},...,f_{\ell,\omega_k}^{(R)} ]'$) over the square 
root of the 
product of their autospectra at $\omega_k$ (see \citet{ShumwayStoffer}). 
Coherency is complex-valued and lies inside the unit circle 
(i.e., its magnitude is less than or equal to $1$). Here, we 
consider \textit{coherence}, denoted by $\kappa_{\ell,\ell'; \omega_k}$,  which 
is the squared modulus of the coherency and thus lies in 
$[0,1]$. 

We consider two approaches to estimating $\kappa_{\ell,\ell'; \omega_k}$.
In the first case, when 
the dependence between microelectrodes $\ell$ and $\ell'$, $\ell,\ell'=1,...,d$, is constant across all epochs, we have 
the estimator
\begin{eqnarray*}
\widehat{\kappa}_{\ell,\ell'; \omega_k}&:=& \frac{ \left| \sum_{r=1}^{R}f_{\ell,\omega_k}^{(r)}f_{\ell',\omega_k}^{\star (r)} \right|^2}
{\sum_{r=1}^{R}\left( f_{\ell,\omega_k}^{(r)}f_{\ell,\omega_k}^{\star (r)} \right) \sum_{s=1}^{R}\left( f_{\ell',\omega_k}^{(s)}f_{\ell',\omega_k}^{\star (s)} \right)}
\end{eqnarray*}
where $f^{\star (r)}$ refers to the complex conjugate of $f^{(r)}$. 
In practice, we are interested in estimating the dependence over a 
{\it band of frequencies} rather than single-valued frequencies. Thus, 
the second case  is justified: we assume that we are interested by the dependence on a fixed epoch $r=1,...,d$; we compute the latter over frequencies (i.e. over the frequencies of a given band in the set $\mathcal{K}:=\{ \Delta:[0,4)\text{Hertz},$ $ \theta:[4,8)\text{ Hertz},$ $ \alpha:[8,12)\text{ Hertz},$ $ \beta:[12,30)\text{ Hertz},$ $ \gamma \geq 30\text{ Hertz}
 \}$). Hence, the estimator is 
\begin{eqnarray*}
\widetilde{\kappa}_{\ell,\ell'; \mathcal{K}_l}^{(r)}&=& \frac{ \left| \sum_{\omega_k \in \mathcal{K}_l}f_{\ell,\omega_k}^{(r)}f_{\ell',\omega_k}^{\star (r)} \right|^2}
{\sum_{\omega_k \in \mathcal{K}_l}\left( f_{\ell,\omega_k}^{(r)}f_{\ell,\omega_k}^{\star (r)} \right) \sum_{\omega_k \in \mathcal{K}_l}
\left( f_{\ell',\omega_k}^{(r)}f_{\ell',\omega_k}^{\star (r)} \right) }
\end{eqnarray*}
where $\mathcal{K}_l$, $l=1,...,5$ represents a particular frequency band. We remark that indeed, the cardinality of $\mathcal{K}_l$ depends on how big are $T$ is. Thus, for the rest of this article, we consider the first case. One remarks that these estimators are analog versions, for the spectral domain analysis, of the sample estimation of Pearson's correlation matrix. Thus, we make the following assumption.
\begin{assu}
\label{assuCoherence}
(Convergence in probability of the coefficient of dependence): The asymptotic behavior of $\widehat{\kappa}_{\ell,\ell'; \omega_k}$ (or of $\widetilde{\kappa}_{\ell,\ell'; \mathcal{K}_l}^{(r)}$) relative to $\kappa_{\ell,\ell'; \omega_k}$ is analog to the one of the sample Pearson's correlation relative to the real Pearson's correlation measure. Hence, if the second joint moment between the variables on which the correlation is measured is finite (e.g $\mathbb{E}\left[ (\delta_{\ell,\omega_k})^2(\delta_{\ell',\omega_k})^2\right] < \infty$), then $\widehat{\kappa}_{\ell,\ell'; \omega_k} \xrightarrow{P} \kappa_{\ell,\ell'; \omega_k}$ as $R \to \infty$. 
\end{assu}

\begin{prop}
(Convergence in probability of the copula model): Let $\xi_{\ell,\omega_k}$ and $\xi_{\ell',\omega_k}$ be two continuous cumulative distributions functions, $\theta$ (properly denoted $\theta_{\ell,\ell';\omega_k}$ that we simplified on purpose) be a monotone function of $\kappa_{\ell,\ell'; \omega_k}$, $\widehat{\theta}$ be the same monotone function applied to $\widehat{\kappa}_{\ell,\ell'; \omega_k}$; and $u,v$ two uniform observations lying on the unit interval. Let $C_\theta$ be the parametric copula model of $C_{\ell,\ell';\omega_k}$. Thus, 
$C_{\widehat{\theta}} (\hat{\xi}_{\ell,\omega_k}(u),\hat{\xi}_{\ell',\omega_k}(v)) \xrightarrow{P}  C_\theta (\xi_{\ell,\omega_k}(u),\xi_{\ell',\omega_k}(v))$ when $R \to \infty$.
\end{prop}

\noindent \textit{Justification:} 
Using the Glivenko-Cantelli Lemma, it is obvious that $C_\theta (\widehat{\xi}_{\ell,\omega_k}(u),\widehat{\xi}_{\ell',\omega_k}(v))$ converges almost surely to $C_\theta (\xi_{\ell,\omega_k}(u),\xi_{\ell',\omega_k}(v))$. Thus, it implies that $C_\theta (\widehat{\xi}_{\ell,\omega_k}(u),\widehat{\xi}_{\ell',\omega_k}(v))$ converges in probability to $C_\theta (\xi_{\ell,\omega_k}(u),\xi_{\ell',\omega_k}(v))$. Now, we have to show that $C_{\widehat{\theta}} (\widehat{\xi}_{\ell,\omega_k}(u),\widehat{\xi}_{\ell',\omega_k}(v))$ converges in probability to $C_\theta (\widehat{\xi}_{\ell,\omega_k}(u),\widehat{\xi}_{\ell',\omega_k}(v))$. In other words, we must show that $\lim_{R \to \infty} \mathbb{P}\left(   \left| \widehat{\kappa}_{\ell,\ell'; \omega_k}-\kappa_{\ell,\ell'; \omega_k} \right| > \epsilon \right)=0$ $ \forall\epsilon >0$, which is the same than showing that $\lim_{R \to \infty} \mathbb{P}\left(   \left| \widehat{\theta}-\theta \right| > \epsilon \right)=0 \forall \epsilon >0$ due to the fact that $\theta$ is a one-to-one function of $\kappa_{\ell,\ell'; \omega_k} $. It is directly demonstrated due to Assumption \ref{assuCoherence}.\newline
\subsubsection{Kendall's rank-based coherence}
Since coherence measures only linear associations between a pair of signals, 
it is important to look into other approaches that could express
dependence via non-linear measures that can be used in inference of 
association parameters for non-elliptical copulas. For this reason, we 
consider rank-based dependence measures; especially because their direct relations with Archimedean copulas are well studied in the literature. We propose here a rank-based coherence measure which is the direct 
analogue of Kendall's tau applied the spectral domain. To the best 
of our knowledge, such a nonparametric measure of rank correlation has never been studied nor proposed before. We note that the same approach for a rank correlation in the sense of Spearman's rho is possible. However, for certain copula families, Kendall's approach leads to a closed analytic form of the copula dependence parameter while Spearman's approach 
does not.

Let the rank-based coherence being computed over epochs $R$, between channels $\ell$ and $\ell'$; $\ell,\ell'=1,...,d$. Hence, one estimates $\mathbb{K}_{\ell,\ell';\omega_k}$ by $\widehat{\mathbb{K}}_{\ell,\ell';\omega_k}=\frac{\mathcal{C}_{\ell,\ell';\omega_k} - \mathcal{D}_{\ell,\ell';\omega_k}}{R(R-1)/2}$ where 
{\small
\begin{eqnarray*} 
\mathcal{C}_{\ell,\ell';\omega_k}&=&\sum_{r=1}^{R} \sum_{s=1}^R \sum_{t=r}^{R} \sum_{w=s+1}^{R} 
\left[\mathbb{I}\left( \delta_{\ell,\omega_k}^{(r)} < \delta_{\ell',\omega_k}^{(s)}, \delta_{\ell,\omega_k}^{(t)} < \delta_{\ell',\omega_k}^{(w)}\right)
+\mathbb{I}\left( \delta_{\ell,\omega_k}^{(r)} > \delta_{\ell',\omega_k}^{(s)}, \delta_{\ell,\omega_k}^{(t)} > \delta_{\ell',\omega_k}^{(w)}\right) \right] \\
 & + & 
\sum_{r=1}^{R} \sum_{s=1}^R \sum_{t=r+1}^{R} \sum_{w=1}^{s} 
\left[\mathbb{I}\left( \delta_{\ell,\omega_k}^{(r)} < \delta_{\ell',\omega_k}^{(s)}, \delta_{\ell,\omega_k}^{(t)} < \delta_{\ell',\omega_k}^{(w)}\right)
+\mathbb{I}\left( \delta_{\ell,\omega_k}^{(r)} > \delta_{\ell',\omega_k}^{(s)}, \delta_{\ell,\omega_k}^{(t)} > \delta_{\ell',\omega_k}^{(w)}\right) \right],
\end{eqnarray*}
\begin{eqnarray*}
\mathcal{D}_{\ell,\ell';\omega_k} &=& \sum_{r=1}^{R} \sum_{s=1}^R \sum_{t=r}^{R} \sum_{w=s+1}^{R} 
\left[\mathbb{I}\left( \delta_{\ell,\omega_k}^{(r)} < \delta_{\ell'\omega_k}^{(s)}, \delta_{\ell,\omega_k}^{(t)} > \delta_{\ell',\omega_k}^{(w)}\right)
+\mathbb{I}\left( \delta_{\ell,\omega_k}^{(r)} > \delta_{\ell',\omega_k}^{(s)}, \delta_{\ell,\omega_k}^{(t)} < \delta_{\ell',\omega_k}^{(w)}\right) \right]\\
 &+& \sum_{r=1}^{R} \sum_{s=1}^R \sum_{t=r+1}^{R} \sum_{w=1}^{s} 
\left[\mathbb{I}\left( \delta_{\ell,\omega_k}^{(r)} < \delta_{\ell'\omega_k}^{(s)}, \delta_{\ell,\omega_k}^{(t)} > \delta_{\ell',\omega_k}^{(w)}\right)
+\mathbb{I}\left( \delta_{\ell,\omega_k}^{(r)} > \delta_{\ell',\omega_k}^{(s)}, \delta_{\ell,\omega_k}^{(t)} < \delta_{\ell',\omega_k}^{(w)}\right) \right]
\end{eqnarray*}}
are the concordances and discordances, respectively. To infer for a single epoch, conditioning on frequencies (or over a specific band), the estimation is, analogously to the coherence measure, a summation over the frequencies of a given band $\mathcal{K}_l$ instead of over many epochs. It is interesting to note that, following similar arguments than for coherence measure, the asymptotic convergence of $\hat{\mathbb{K}}_{\ell,\ell';\omega_k}$ may be shown based on the original work of \citet{kendall1948}.

\subsection{Model-examples}
\subsubsection{Elliptical family}
We extend our model to the field of elliptical copulas (see \citet{nelsen2007}), by applying to this well-known family the spectral-based elements discussed above. One of the main advantages in the use of elliptical copulas is related to the dependence parameter: for this family, the dependence is expressed as the Pearson's correlation matrix. With the analogy between coherence and correlation, moving to the spectral domain, we could directly use the coherence measure as the Fourier transforms are expressed in term of periodograms, similarly to the coherency expressed in term of coherence.

One observation here is that due to the range of the coherence measure $[0,1]$, we have to impose positive dependence only when 
elliptical copulas are used within our proposed framework.
For instance, let $\Phi$ the standard cdf of a Gaussian distribution and $\Phi^{-1}$ its inverse. Let $u=\widehat{\xi}_{\ell,\omega_k}(z_1)$ and $v=\widehat{\xi}_{\ell',\omega_k}(z_2)$. Then, the bivariate semi-parametric estimator of the coherence-based Gaussian copula computed on a frequency $\omega_k$ is expressed by 

\begin{eqnarray*}
\widehat{C}_{\widehat{\kappa}_{\ell,\ell'}}^{gaussian}(\hat{\xi}_{\ell,\omega_k}(x_\ell),\hat{\xi}_{\ell',\omega_k}(x_{\ell'})) &=& \int_{-\infty}^{\Phi^{-1}(\hat{\xi}_{\ell,\omega_k}(x_\ell))} \int_{-\infty}^{\Phi^{-1}(\widehat{\xi}_{\ell',\omega_k}(x_{\ell'}))} \frac{1}{2\pi (1-\widehat{\kappa}_{\ell,\ell';\omega_k}^2)^{1/2}} \times \\ & & \hspace{3cm}\exp \left\{ -\frac{s^2-2 \widehat{\kappa}_{\ell,\ell';\omega_k} st +t^2}{2(1-\widehat{\kappa}_{\ell,\ell';\omega_k}^2)} \right\}ds\text{  }dt \\
  &=& \Phi^{(r)}_{\widehat{\kappa}_{\ell,\ell'}^{(r)}}\left( \Phi^{-1}(\widehat{\xi}_{\ell,\omega_k}(x_\ell)),\Phi^{-1}(\widehat{\xi}_{\ell',\omega_k}(x_{\ell'})) \right)
\end{eqnarray*}
where one estimates a single copula using data from all epochs.

\subsubsection{Archimedean family}

With respect to the inference approach for elliptical copulas, the main difference for Archimidean copulas concerns the dependence parameter: instead of working with a straight expression of the coherence as a parameter, we consider functionals of the rank-based coherence (see \citet{genest1993}). In fact, we consider the same functionals as for the case of inference using Kendall's tau. Hence, taking the example for a Clayton copula, one obtains
 \begin{eqnarray*}
\widehat{C}_{\widehat{\theta}_{\ell,\ell'}}^{Clayton}(\hat{\xi}_{\ell,\omega_k}(x_\ell),\hat{\xi}_{\ell',\omega_k}(x_\ell')) &=& \left[(\hat{\xi}_{\ell,\omega_k}(x_\ell))^{-\theta}+(\hat{\xi}_{\ell',\omega_k}(x_{\ell'}))^{-\theta} \right]^{-1/\theta}
\end{eqnarray*}
where, the function linking $\theta$ to $\mathbb{\kappa}_{\ell,\ell'; \omega_k}$ here, is $\mathbb{K}=1+\frac{\theta}{2}$. Again, it is important to reiterate that we estimate the copula using data from all epochs.

\section{Simulations}
\label{simuls}
These simulations are divided in two parts. In the first part, we show that working on a specific frequency instead of on the original data themselves may lead to the detection of strong hidden dependencies, in particular in the spectral domain. Hence, it may improve the robustness of the multivariate function of probability due to the fact that in such a case, dependence is less likely to have been due to just random chance. Also, it could lead to a more robust multivariate function of probability since this is less sensitive to noise. In the second part, we show the our approach captures non-linear dependencies that standard standard linearity-based methods do not. 

\subsection{Simulation 1: Dependencies in the spectral domain instead of in the time domain}\label{SimuSec1}

We show one feature: we illustrate that our methodology catches strong dependence hidden into frequencies. We do that by comparing the rank-based coherence measured over epochs on a single frequency to the Kendall's tau measured  on original data. 

\noindent 

Let $Z_t^{(r)}$ be a latent signal which is an second order autoregressive 
(AR(2)) process $Z_t^{(r)} = 1.989 Z_{t-1}^{(r)} - 0.990 Z_{t-2}^{(r)} + W_t^{(r)}$ 
where $W_t^{(r)}$ is a white noise sequence. Let the sampling rate be $1500 \text{ Hertz}$. The roots of this AR(2) polymomial function are complex-valued with magnitude 
$1.005$ and phase $2 \pi \frac{12}{1500}$ so that the spectra 
of this latent process has power concentrated around the phase. 
The observed time series $X_t^{(r)}$ and $Y_t^{(r)}$ are defined by
$X_t^{(r)} = 0.90 Z_{t-1}^{(r)} + \epsilon_{X,t}^{(r)}; Y_t^{(r)} = 0.85 Z_t^{(r)} + 
\epsilon_{Y,t}^{(r)}$
where $\epsilon_{X,t}^{(r)}$ and $\epsilon_{Y,t}^{(r)}$ are independent 
of each other and each is a white noise with identical variance 
$\sigma_{\epsilon}^2$. There were a total of $2000$ replicated 
datasets. Each dataset consisted of $R=1000$ epochs 
and the total number of observations for each epoch was $T=1500$. 
For each of these $2000$ datasets, we computed a rank-based measure (Kendall's tau) on all data, and we computed our proposed rank-based measure (rank-based coherence) on a single frequency. For this example, rank-based coherence was 
calculated only for frequency $2 \pi \frac{12}{1500}$ which is the location of the peak of the spectrum of $Z_t^{(r)}$.

\begin{figure}[!htbp]
\begin{center}
\includegraphics[scale=0.75 ]{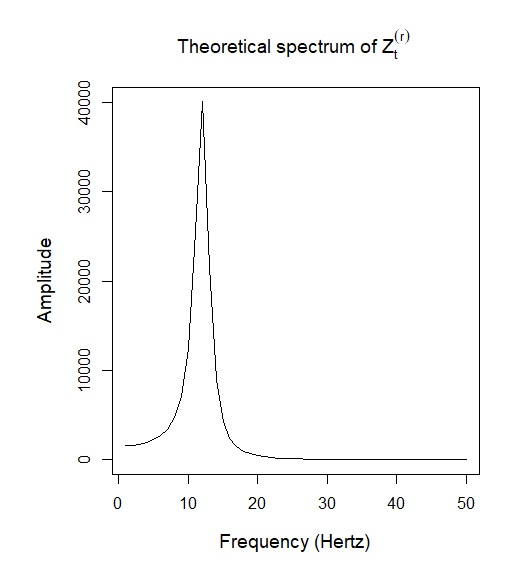}
\caption{\textbf{Left: }Theoretical spectrum of the simulated latent signal in Section \ref{SimuSec1}.\textbf{Right: }Zoom in of that theoretical spectrum, on the interval [0,50] Hertz. }
\label{Sim}
\end{center}
\end{figure}


\begin{table}
\begin{center}
\begin{tabular}{|l||*{5}{c|}}

    \hline
     Dependence & mean & median & variance & minimum & maximum   \\
		 measure    &  & & & &  \\
    \hline \hline
    Kendall's tau & $0.2942$ & $0.2943$ & $7.47 \times 10^{-6}$ & $0.2876$ & $0.3025$ \\
		on original data $(\tau)$ & & & & & \\
		\hline
		Rank-based & $0.8694$ & $0.8707$ & $2.15 \times 10^{-4}$ & $0.8214$ & $0.8998$ \\
		 coherence $(\mathbb{K})$& & & & & \\
		\hline
		  \end{tabular}
	\caption{Top row: Kendall's tau measure based on the original time series. Bottom row: rank-based coherence based on frequency 12 Hertz. }
	\label{tableKendall}
\end{center}
\end{table}

The goal of this first part was to note a difference in the strength of the dependence between the original data, measured by Kendall's tau, where epochs are considered only as segments of the dataset due to the stationarity assumption; and amplitudes of the Fourier transforms on all epochs, measured at 12 Hz, by Kendall rank-based coherence. Results are shown in Table \ref{tableKendall}. Note that under the null scenario of no difference in the dependence, the difference in the distributions of the dependence measures is obvious.

\subsection{Simulation 2: Assessment of non-linear dependencies}
\begin{figure}[!htbp]
\begin{center}
\includegraphics[scale=0.80 ]{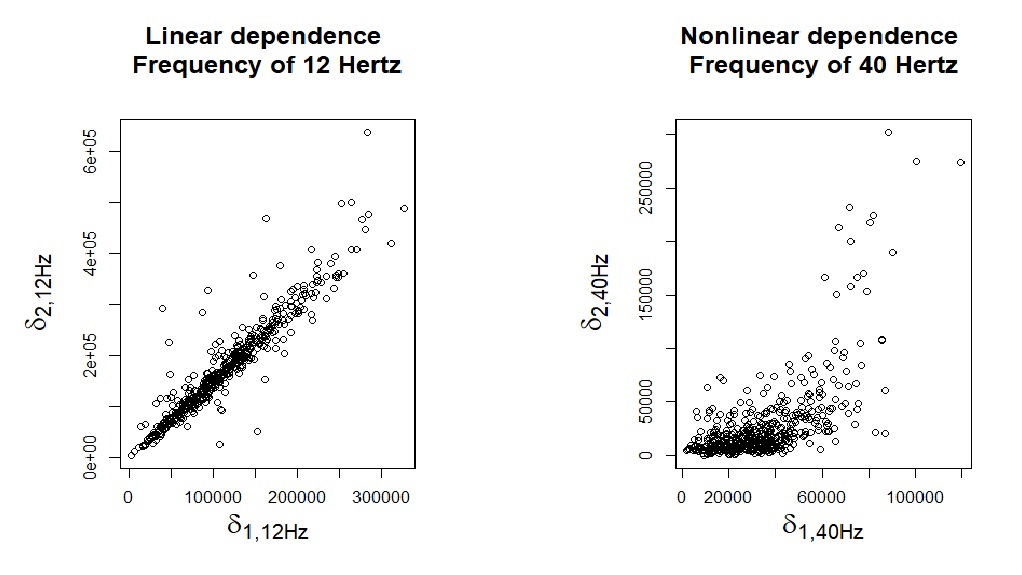}
\caption{\textbf{Left: }Dependence between the Fourier coefficients for a frequency of $12$ Hertz. \textbf{Right: }Dependence for a frequency of $40$ Hertz. The nature of dependence changes between frequency bands. For the alpha band (12 Hertz) the dependence is linear; for the beta band (40 Hertz) the dependence appears not following a linear path. }
\label{Simuls}
\end{center}
\end{figure}

For this part, we generated two types of latent signals from AR(2) processes in such a way that the spectra of these latent signals are concentrated of the phase of the frequencies of interest. The principle here is that we have, for $r=1,...,500$ and for $t=1,...,1000$  the following latent signals: $Z_\alpha^{(r)}(t) \sim AR(2)$ with polynomial function having complex-valued roots of phase $p_\alpha=\pm 12/1000\times 2 \pi$ (which means that the spectra if concentrated around $12$ Hertz), and  $Z_\beta^{(r)}(t) \sim AR(2)$ of phase $p_\beta=\pm 40/1000\times 2 \pi$. From these latent signals, we generated the two following observed signals:
\begin{itemize}
\item $X_1^{(r)}(t)=Z_\alpha^{(r)}(t)+Z_\beta^{(r)}(t)+\epsilon_t^{(r)}$
\item $X_2^{(r)}(t)=\frac{3}{2}Z_\alpha^{(r)}(t)+\eta \left( Z_\beta^{(r)}(t) \right)^4  \sin( Z_\beta^{(r)}(t))  + \epsilon_t^{(r)}$
\end{itemize}
where $\eta = 1/100000$,$r=1,...,500$, $t=1,...,1000$ and $\epsilon_t^{(r)}\sim \mathcal{N}(0,0.01\mathbb{V}(Z_1\beta^{(r)}(t))$. We replicated this simulation scenario $B=10000$ times. We denote $\delta_{\ell, \omega_k}=[\delta_{\ell, \omega_k}^{(1)},...,\delta_{\ell, \omega_k}^{(500)}]'$, $\ell=1,2$; the vector of length $r=500$ of the Fourier coefficients for a fundamental frequency $\omega_k$, applied on observed signal $X_\ell^{(r)}(t)$. Thus, in Figure \ref{Simuls}, one sees the dependence for the Fourier coefficients filtered at $12$ and $40$ Hertz between the two observed signals. One notices that obviously a linearity-based dependence measure is suitable for the filtering at $12$ Hertz. However, in the case of the filtering at $40$ Hertz, any linearity-based dependence measure will fail to catch this non-linear dependence, while a well-specified copula will. We note that we selected the copula model for these simulations based on AIC \cite{akaike1987}. Indeed, \citet{jordanger2014}  have shown that AIC, based on a penalized likelihood approach on the number of parameters of the model, has almost no difference in the choice of a model with other copula-based information criterions when copula are estimated on large populations.

Table \ref{tableSimuls} shows the results of these simulations by exhibiting the mean of $\mathbb{K}_{1,2;\omega_k}^b$ (which denotes the Kendall-based coherence measure for Simulation $b=1,...,10000$ at frequency $\omega_k$ between variables $X_1$ and $X_2$), the most selected copula model among all the simulations and it's inherent dependence parameter. Table \ref{tableFreqSim} shows the dispersion of the selected copula model for the $10000$ simulations.

\begin{table}[!htbp]
\begin{center}
\begin{tabular}{|l||*{2}{c|}}

    \hline
     Fundamental & $\bar{\mathbb{K}}_{1,2; \omega_k}^b$ & Selected copula model    \\
		 frequency    &  & $\bar{\hat{\theta}}$  \\
    \hline \hline
    $12$ Hertz & $0.8397$ & Frank  \\
		           &          & $23.45$  \\
		\hline
		$40$ Hertz & $0.4172$ & Gumbel   \\
		           &          &  $1.65$        \\
		\hline
		  \end{tabular}
	\caption{Averaged Kendall-based coherence measure, selected copula model and parameter of copula related to the selected copula model, over 10 000 simulations. }
	\label{tableSimuls}
\end{center}
\end{table}

\begin{table}[!htbp]
\begin{center}
\begin{tabular}{|l||*{7}{c|}}

    \hline
     Fundamental & Independent & Gaussian & Student & Clayton & Gumbel & Frank & Joe    \\
		 frequency    &  & & & & &  &  \\
    \hline \hline
    $12$ Hertz & 0& 0 & 0& 0& 27 & 9973 & 0  \\
		\hline
		$40$ Hertz & 0 & 63 & 0 & 0 & 9628 & 0 & 309   \\
		\hline
		  \end{tabular}
	\caption{Frequency of selection, for each frequency band, between $\delta_{1,\omega_k}$ and ${\delta}_{2,\omega_k}$, over the 10 000 simulations }
	\label{tableFreqSim}
\end{center}
\end{table}

\section{Application: Local field potential of rats} \label{lfp}

We apply our copula-based approach for modeling dependence in local field potential (LFP) of rats based on the experimentation of Frostig and \citet{wann2017}. Microelectrodes were inserted in $32$ locations on the rat cortex (4 layers, respectively at $300 \mu m,$ $700\mu m,$ $1100 \mu m$ and $1500\mu m$; 8 microelectodes lined up in each layer). From these microelectrodes, $T=1000$ time points were recorded per second. As we assume a stationary behavior within each second, we consider each second as a distinct epoch $r$. A total of $r=600$ epochs were recorded. Midway in this period (at epoch $r=300$), stroke was mechanically induced on each rat. 

In this paper, as the scope is not about assessing a difference between two different copulas, but is rather to write adequately these copulas with respect to the idiosyncrasies of data, we limit ourselves here to show the copula-based modeling done on LFP data. Thus, we present three different situations which might lead to a further analysis. We remind the reader that the selection of a model of copula is not the topic of this paper; hence we based our selection on a well-accepted criterion in the copula literature: AIC (\citet{akaike1987}); among the following copula models: Independent, Gaussian, Student, Clayton, Gumbel, Frank and Joe.

\subsection{Modeling the copula between two microelectrodes for a given frequency} 
We are interested here by modeling the dependence between two different microelectrodes, for a frequency of $12$ Hertz, for the whole course of the experiment (data from the LFP recording are considered for the entire $600$ seconds). In this case, because the nature of the dependence between electrodes is intrinsic to each rat, we apply our method to only one rat: rat id $141020$.
\subsubsection{First case: Highly-dependent microelectrodes}
\begin{figure}[!htbp]
\begin{center}
\includegraphics[scale=0.75 ]{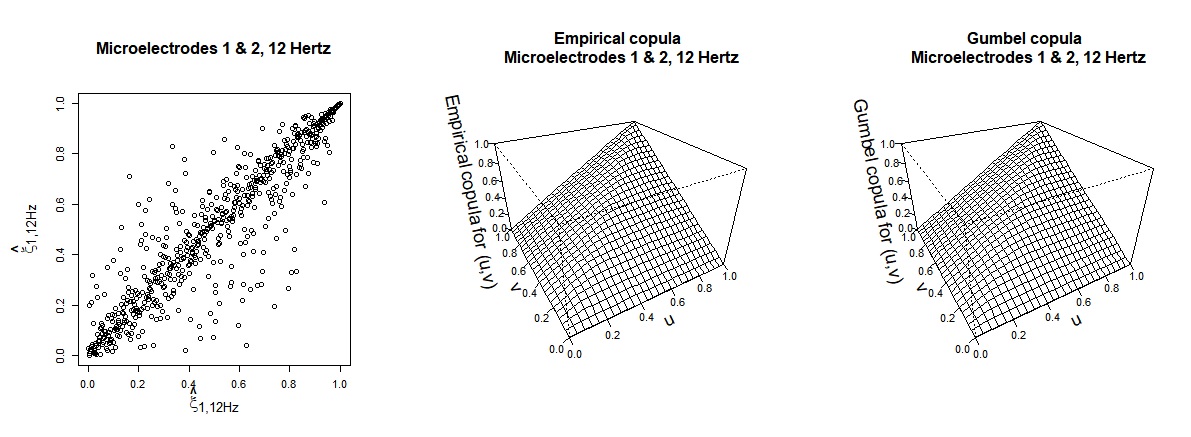}
\caption{\textbf{Left:} Scatterplot of the dependence between the empirical cdf for microelectrodes $1$ and $2$, at a frequency of $12$ Hertz. \textbf{Middle:} Empirical copula between microelectrodes $1$ and $2$. \textbf{Right:} Joe copula between both microelectrodes. }
\label{lfp1_2}
\end{center}
\end{figure}

\begin{figure}[!htbp]
\begin{center}
\includegraphics[scale=0.35 ]{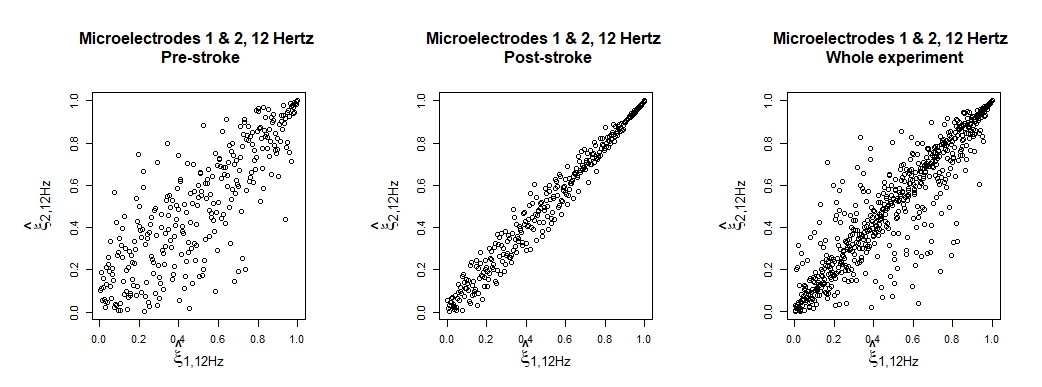}
\caption{\textbf{Left:} Scatterplot of the dependence $\delta_{1, 12Hz}$ and $\delta_{2, 12Hz}$ for the $300$ epochs pre-stroke. \textbf{Middle:} Dependence  between $\delta_{1, 12Hz}$ and $\delta_{2, 12Hz}$ for the $300$ epochs post-stroke. \textbf{Right:} Dependence  between $\delta_{1, 12Hz}$ and $\delta_{2, 12Hz}$ for the whole experiment (600 epochs). }
\label{lfp1_2_prepost}
\end{center}
\end{figure}

We consider the case where dependence between two microelectrodes (channels) is high, from Kendall rank-based coherence perspective (at the alpha frequency band). Thus, we considered dependence between microelectrodes $1$ and $2$ (two microelectrodes on columns $1$ and $2$ of the first layer) where $\mathbb{K}_{1,2; 12Hz}=0.753$. Figure \ref{lfp1_2} shows (left) the dependence between the empirical cdf of both microelectrodes. Based on Akaike information criterion \cite{akaike1987}, this dependence is well represented through a Gumbel copula (right) of parameter $4.12$. We also included a plot of the empirical copula (middle); for more about the empirical copula, see \citet{deheuvels1979} in order to show that visually, the empirical copula is close to the one chosen from AIC. We note that in Figure \ref{lfp1_2_prepost}, this dependence is graphically represented for the pre-stroke period (epochs $r=1,...,300$) as well as for the post-stroke period (epochs $r=301,...,600$).

\begin{figure}[!htbp]
\begin{center}
\includegraphics[scale=0.35 ]{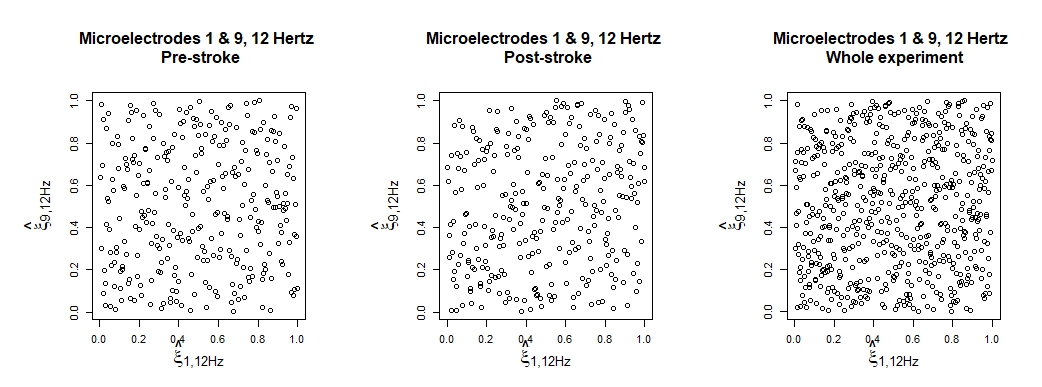}
\caption{\textbf{Left:} Scatterplot of the dependence $\delta_{1, 12Hz}$ and $\delta_{9, 12Hz}$ for the $300$ epochs pre-stroke. \textbf{Middle:} Dependence  between $\delta_{1, 12Hz}$ and $\delta_{9, 12Hz}$ for the $300$ epochs post-stroke. \textbf{Right:} Dependence  between $\delta_{1, 12Hz}$ and $\delta_{9, 12Hz}$ for the whole experiment (600 epochs). }
\label{lfp1_9_prepost}
\end{center}
\end{figure}

\subsubsection{Second case: Independent microelectrodes}\label{ind}
\begin{figure}[!htbp]
\begin{center}
\includegraphics[scale=0.75 ]{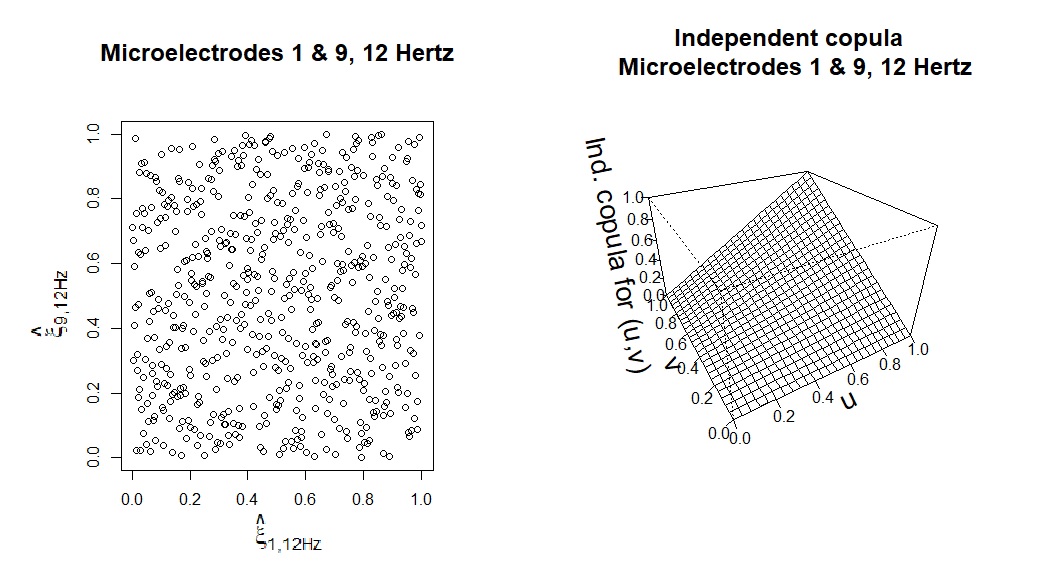}
\caption{\textbf{Left:} Scatterplot of the dependence between the empirical cdf for microelectrodes $1$ and $9$, at a frequency of $12$ Hertz.  \textbf{Right:} Independence copula between the two microelectrodes. }
\label{lfp1_9}
\end{center}
\end{figure}

Still working with microelectrode $1$, we decided to model the dependence of the latter with his other neighbor (on the same column): microelectrode $9$. Thus, we computed the Kendall rank-based coherence and obtained a value of $\mathbb{K}_{1,9; 12Hz}=0.041$; which is small enough to suggests no strong evidence to conclude dependence between microelectrodes. Thus, we applied the Kendall's based independence test for bivariate samples (for more about this test, see \citet{genest2007}) where the null hypothesis is the independence between the microelectrodes and the test statistic is 

$$\text{\textit{Test statistic}}:= \mathbb{K}_{1,9; 12Hz}|\sqrt{\frac{9R(R-1)}{2(2R+5)}}$$
where $R=600$ is the number of epochs. The \textit{p-value} for this test is 0.13 which indicates insufficient evidence against independence. Thus, the independence copula (simply the product of the margin) is the adequate choice here. In Figure \ref{lfp1_9}, one observes that independence between in the scatterplot (left) and the related copula (right), which is no more than the product of the margins. In this case again, the dependence for the pre-stroke period as well as the one for the post-stroke period are shown in Figure \ref{lfp1_9_prepost}.

\subsection{Modeling the copula between an epoch during pre-stroke and an epoch during post-stroke}
\begin{figure}[!htbp]
\begin{center}
\includegraphics[scale=0.75 ]{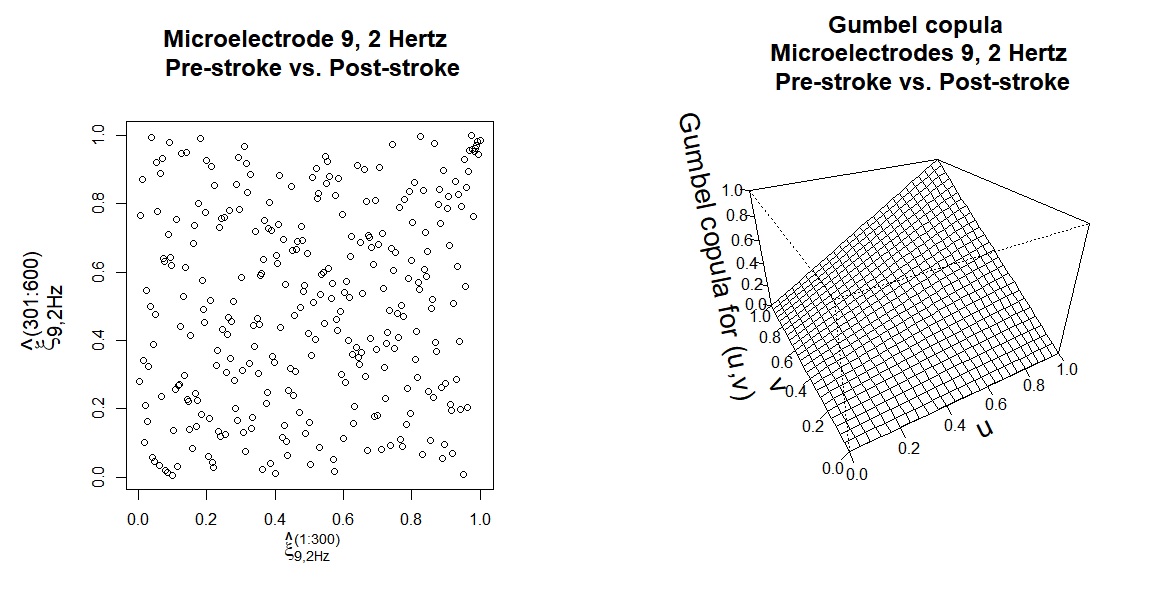}
\caption{\textbf{Left:} Scatterplot of the dependence between the empirical cdf of pre-stroke versus post-stroke, for microelectrode $9$ at a frequency of $2$ Hertz.  \textbf{Right:} Independence copula between the two microelectrodes. }
\label{lfp9}
\end{center}
\end{figure}

The principle here is to model the dependence structure, for a given Fourier fundamental frequency and a given microelectrode, between the pre-stroke period and the post-stroke's one. As the rat's brain activity is perturbed by the induced stroke, for many microelectrodes and for many frequencies, one observes quasi-independence. Hence, we present here the case for a very small dependence, but where the independence test (described in \ref{ind}) applied between $\delta_{9,2Hz}^{1:300}$ and $\delta_{9,2Hz}^{301:600}$, gives a \textit{p-value}$<0.05$. We introduce the superscripts for the epochs $(s:t), (s':t'), s<t, s'<t'$ in the notation $\mathbb{K}_{\ell,\ell';\omega_k}^{(r:s),(r':s')}$ where that superscript has a similar meaning than for $\delta_{\ell,\omega_k}^{(r:s)}$: for $\delta_{\ell,\omega_k}$, microelectrode $\ell$ is measured respectively from epochs $r$ to $s$ and from epochs $r'$ to $s'$. Thus, we considered microelectrode $9$ for the simple reason that it has been difficult to find a microelectrode, for a given frequency, exhibithing where the independence test does not indicate the independence between the epochs pre-stroke and those post-stroke. Then, for a frequency of $2$ Hertz, the Kendall's rank-based coherence between $\delta_{9,2Hz}^{(1:300)}$ and $\delta_{9,2Hz}^{(301:600)}$ is $\mathbb{K}_{9,9; 2Hz}^{(1:300)(301:600)}=0.116$. Thus, using AIC, we selected a Gumbel copula of parameter $1.113$. Figure \ref{lfp9} shows this weak dependence (left) where one observes a little upper tail dependence. On the right, one observed the perspective of the fitted copula.

\section{Discussion}
\label{conclusion}
This paper proposed a new approach to express dependence between two time series at a given frequency using copulas in the spectral domain. We provided the necessary methodological framework and proposed a rank-based coherence, strongly inspired by Kendall's tau measure, in order to infer adequately a semi-parametric copula function (parametric copula model with non-parametric margins) to deal with data represented in the spectral-domain. Our simulations show that 
the copulas parametrized between two raw time series (i.e., time domain 
data) and the one parametrized with modulus of the Fourier transform 
on specific frequencies (i.e., spectral domain data) might be totally two different objects. Crucially, even when dependence between two time series appears to be weak, we can build particular probabilities functions (i.e. copulas) on specific frequencies of their spectrum; therefore we are able to model stronger dependence. Finally, we illustrate the applicability of our methodology through the local field potential of some rats.

This work opens the way to model, for example, dependence between channels (represented through microelectrodes in this paper) of local field potential which will potentially lead to a better understanding of brain connectivity. Non-linear dependence is not a rare phenomenon with neuroimaging data represented in the spectral domain. However, very few papers have dealt with this specific problem and thus our hope is that this paper can make some contribution on this front.  

\section*{Acknowledgements}
Ron D. Frostig was supported by the Leducq Foundation (grant 15CVD02).
\bibliographystyle{chicago}

\bibliography{refSPL}

\end{document}